\documentclass{aastex}
\usepackage{spr-astr-addons}

\RequirePackage{color}

\usepackage{latexsym}
\usepackage{graphics}
\usepackage{graphicx}
\usepackage{epsfig}

\begin{document}

\title{ Charge-changing transitions and capture strengths of pf-shell nuclei with $T_{z} = -
2$ at proton drip-line } \shorttitle{Capture rates and proton
drip-line nuclei}
\author{Muneeb-Ur Rahman\altaffilmark{1}}
\affil{Department of Physics, Islamia College Peshawar, KP,
Pakistan\\ email: muneeb@icp.edu.pk}
\author{Jameel-Un Nabi\altaffilmark{2}}
\affil{Faculty of Engineering Sciences, GIK Institute of Engineering
Sciences and Technology, Topi 23640, Swabi, KP, Pakistan\\ email:
jameel@giki.edu.pk }

\begin{abstract}
Charge-changing transitions, commonly referred to as Gamow-Teller
(GT) transitions, and electron capture/$\beta^{+}$-decay strengths
for pf-shell nuclei with $T_{z} = - 2$ at proton drip-line have been
calculated using the proton-neutron quasi-particle random phase
approximation (pn-QRPA) model. The total GT$_{+}$ strength values
and electron capture/$\beta^{+}$-decay rates are needed for the
study of the late stages of the stellar evolution. The pn-QRPA
theory is used for a microscopic calculation of GT strength
distribution functions and associated stellar electron
capture/$\beta^{+}$-decay rates of proton-rich pf-shell nuclei with
$T_{z} = - 2$ in the mass range 46 $\leq$ A $\leq$ 56 at proton
drip-line. Standard quenching factor of 0.74, usually implemented in
the shell model calculation, has been incorporated for the
comparison with experimental data (wherever available).  The
calculated GT strength of the two proton-rich nuclei, $^{52}$Ni and
$^{56}$Zn are compared with experimental data of corresponding
mirror nuclei. It has been found that the pn-QRPA results are in
good agreement with the experimental data as well as shell model
result.  It is noted that the total GT strength increases linearly
with the increase of mass number. The electron
capture/$\beta^{+}$-decay rates for proton-rich nuclei are
calculated on a temperature and density scale relevant to
presupernova evolution of massive stars. The $\beta^{+}$ decay
half-lives are compared with measured and other theoretical
calculations.
\end{abstract}

\section{Introduction}
The simulation of supernovae explosion mechanism depends on many
input parameters to be fed in relevant mega codes. Alone the
parameters related to nuclear physics involve data for a large
number of nuclei required to simulate this complex scenario. The
more susceptible nuclei for electron capture and beta decay tend to
be small in numbers, however, weak rate times abundance is the
quantity important for necessary action. Synthesis of the iron group
and other heavier elements substantially depend on many input
parameters. Number of electrons per baryon ($Y_{e}$), energetics of
the shock waves, entropy of the stellar core, mass and metalicity of
the progenitor, mixing and fallback, and explosion energy are among
the few. The hydrodynamic shock is believed to be formed at the edge
of homologous core and its energy is related to $Y_{e}$ as:
\begin{eqnarray}
E_{S}\simeq (GM_{HC}^2/R_{HC})(Y_{ef}- Y_{ei})\\
\nonumber\simeq M_{HC}^{5/3}(Y_{ef}-Y_{ei}) \simeq
Y_{ef}^{10/3}(Y_{ef}-Y_{ei}),
\end{eqnarray}
where $M_{HC}$, $R_{HC}$, $Y_{ei}$, and $Y_{ef}$ are mass of the
unshocked inner core (the homologous core), radius of homologous
core, and initial and final lepton fraction, respectively (Kar et
al. 1994; Nabi et al. 2004; Rahman et al. 2014). The central part of
the massive stars ($M \geq 10 M_{\odot}$) consists of iron core
which grows with the passage of time. When this core exceed the
appropriate Chandrasekhar mass limit, the implosion ensues. This
collapse is subsonic and homologous in the inner region of the core
and is supersonic in the outer regions. The behavior of the
supernovae after collapse is very sensitive to the mass of
homologous core, and consequently to final lepton fraction $Y_{ef}$
which in turn is dependent on the electron capture (Bethe 1990). The
mass of the inner core has important consequences such as it sets
the value of kinetic energy imparted to the shock wave, mass cut for
material which the shock has to plough, and sets the amount of
matter and angular momentum that can be dynamically relevant in the
astrophysical scenario. For core density greater than $10^{9}
g/cm^{3}$ the electromagnetic radiation and heat conduction are
transported very slowly as compared to the time scale of the
collapse (Juodagalvis 2010). Thus, for stellar core densities $\leq
10^{11} g/cm^{3}$, provided the stars are not too massive, neutrinos
bleed away from the surface to support low entropy condition and
keep the nucleons bound inside nuclei in stellar interior. This
reduction in entropy in stellar interior favors smaller mass of the
iron core and consequently facilitate the shock's outward march
(Timmes et al. 1996). Lower entropy of the core favors the explosion
because less energy is stored in the nuclear excited states in the
collapsing core and presupernova environment and consequently the
implosion process leads to higher density to produce stronger bounce
and an energetic shock (Bethe et al. 1979). Bethe and collaborators
(Bethe 1990; Bethe et al. 1979) pointed out that the lower central
entropy makes it sensitive to the Chandrasekhar mass limit and due
to the overlying matter's pressure the mass of the final collapsing
core is small as compare to the Chandrasekhar mass. The structure of
the presupernova core (which determines the extent of the convective
shells) and nucleosynthesis in stars are greatly effected by the
entropy profile. Neutrino bleeding, entropy profile and electron to
baryon ratio ($Y_{e}$) are dependent on weak decay rates in stellar
matter. Electron capture reduces the value of $Y_{e}$ and electron
degeneracy pressure in the core to accelerate the collapse. Various
authors (Rahman et al. 2014; Juodagalvis et al. 2010; Heger et al.
2001; Nabi et al. 1999a; Liu 2013; Rahman et al. 2013) implemented
different models to calculate the weak decay rates at temperature
and density scale relevant to astrophysical environment. These
authors noted that the electron capture rates are substantially
suppressed in stellar core as compared to the seminal work of
Fuller, Fowler and Newman (FFN)(Fuller et al. 1980, 1982a, 1982b,
1985). After recognition of the pivotal role of Gamow-Teller (GT)
strength functions in astrophysical environment, FFN used the
parameterization based on the independent particle model to
calculate the GT contributions to the stellar rates. They inserted
the experimental data, available at that time, for the discrete
transitions and assigned a value of \textit{log ft = 5} to
unmeasured allowed GT transitions. However, Caurier and
collaborators (Caurier et al. 1999) observed that for even-even
nuclei their GT centroid was at lower excitation energy in the
daughter nucleus as compared to the FFN. It has also been noted that
for odd-A and odd-odd nuclei, FFN placed GT centroid at too low
excitation energies than the shell model and experimental data
(Aufderheide et al. 1996; Langanke et al. 1998; Nabi and Rahman
2005). Experimental measurements showed that, contrary to the
independent particle model, the total GT strength both in GT$_{+}$
and GT$_{-}$ direction is quenched and fragmented over many final
states in the daughter nucleus (Rapaport et al. 1983; Vetterli et
al. 1989; Anderson et al. 1990; Ronnqvist et al. 1993; El-Kateb et
al. 1994). This quenching and fragmentation of the GT strength is
due to the residual interaction among the valence nucleons and an
accurate description of these correlations is essential and play
significant role for the calculation of stellar weak decay rates.
Various models have been proposed and can be found in literature
elsewhere.

The pn-QRPA is an efficient way to calculate GT strength and
associated weak decay rates (Nabi et al. 1999a; Nabi 2009; Nabi
2012; Nabi et al. 2013). The pn-QRPA model has access to a liberal
model space of 7$\hbar\omega$ to perform the required calculation.
For isospin symmetry, Fermi transitions are simple and they are only
important for $\beta^{+}$ decay of proton-rich nuclei with Z $>$ N
(Sarriguren 2013). The authors (Cole et al. 2012) used QRPA to
calculate and compare GT strength and electron capture rates with
experimental and other models for various pf-shell nuclei and
discussed the pros and cons of their results for astrophysical
scenario. The following section shows the necessary mathematical
formulae for the calculation of GT strength distributions and weak
decay rate at density and temperature scale that are relevant for
astrophysical environment. The GT strength distributions, calculated
in the present study, are compared with shell model results and
measurements in Section~3. The electron capture rates for few
selected proton drip-line nuclei ($^{46}$Mn, $^{48}$Fe, $^{50}$Co,
$^{52}$Ni and $^{56}$Zn) are presented in Section 4. Finally we
conclude our study in Section 5.

\section{Formalism}
For the calculation of GT strength distributions and electron
capture/$\beta^{+}$-decay  rates on proton rich nuclei in stellar
environment, the following main assumptions were taken into account:

1) Only allowed GT and superallowed Fermi transitions were
calculated in the present study. Forbidden transitions are
relatively negligible for the density and temperature scales
considered here.

2) It was assumed that the gas in the stellar medium is completely
ionized and electrons are no longer bound to the nucleus and obey
Fermi-Dirac distribution.

3) Neutrino and anti-neutrino captures escape freely from the
interior and surface of the star for density scales considered in
this project.

4) The effect of particle emission from the excited states were
taken into account.

5) All excited states having energy less than $S_{p}$ ($S_{n}$)
(separation energy of protons (neutrons)) were assumed to decay
directly to the ground state through $\gamma$ transitions.

The Hamiltonian, in the present study, takes the form
\begin{equation} \label{GrindEQ__2_}
{\rm H}^{{\rm QRPA}} {\rm \; =\; H}^{{\rm sp}} {\rm \; +\; V}^{{\rm
pair}} {\rm \; +\; V}_{{\rm GT}}^{{\rm ph}} {\rm \; +\; V}_{{\rm
GT}}^{{\rm pp}},
\end{equation}
where $H^{sp} $, $V^{pair} $, $V_{GT}^{ph} $, $V_{GT}^{pp} $ are the
single-particle Hamiltonian, the pairing force (pairing was treated
in the BCS approximation), the particle-hole (ph) GT force, and the
particle-particle (pp) GT force, respectively. Single particle
energies and wave function were calculated in the Nilsson model,
which takes into account nuclear deformations. The proton-neutron
residual interactions occurred as particle-particle and
particle-hole interaction. The interactions were given separable
form and were characterized by two interaction constants $\kappa$
and $\chi$, respectively. The details and fine tuning of these GT
strength parameters can be found in literature (Staudt et al. 1990;
Muto et al. 1992; Hirsch et al. 1993). Other parameters such as
Nilsson potential parameters (taken from (Nilsson 1955)), the
deformations, pairing gaps, and Q-value of the reactions were used
in the calculation of weak decay rates. Nilsson oscillator constant
was taken as $\hbar\omega = 41A^{-1/3}$ (MeV) for both neutrons and
protons. The traditional choice of $\Delta _{p} =\Delta _{n}
=12/\sqrt{A} (MeV)$ was used for the pairing gaps. Deformation of
the nuclei was calculated as
\begin{equation}
\delta = \frac{125(Q_{2})}{1.44 (Z) (A)^{2/3}},
\end{equation}
where $Z$ and $A$ are the atomic and mass numbers, respectively and
$Q_{2}$ is the electric quadrupole moment taken from Ref.
(M\"{o}ller et al. 1981). Q-value of the reaction was taken from the
mass compilation of Audi and Wapstra (Audi et al. 2003).

The electron capture (ec) and positron decay (pd) rates of a
transition from the $i^{th}$ state of the parent to the $j^{th}$
state of the daughter nucleus are given by
\begin{eqnarray}\scriptsize
\lambda ^{^{ec(pd)} } _{ij} =\left[\frac{\ln 2}{D}
\right]\left[B(F)_{ij} +\left({\raise0.7ex\hbox{$ g_{A}
$}\!\mathord{\left/ {\vphantom {g_{A}  g_{V} }} \right.
\kern-\nulldelimiterspace}\!\lower0.7ex\hbox{$ g_{V}  $}}
\right)^{2} B(GT)_{ij} \right] \nonumber \\
\left[f_{ij}^{ec(pd)} (T,\rho Y_{e}  ,E_{f} )\right]. \label{phase
space}
\end{eqnarray}

The value of D was taken to be 6146 $\pm$ 6 s adopted from Ref.
(Jokinen et al. 2002). Here $B(F)_{ij}$ and $B(GT)_{ij}$ are the
reduced transition probabilities due to Fermi and GT transitions:
\begin{equation}
B(F)_{ij} = \frac{1}{2J_{i}+1} \mid<j \parallel \sum_{k}t_{\pm}^{k}
\parallel i> \mid ^{2}.
\end{equation}
\begin{equation}
B(F)_{ij} = [T(T+1) - T_{zi}T_{zf}].
\end{equation}
\begin{equation}
B(GT)_{ij} = \frac{1}{2J_{i}+1} \mid <j \parallel
\sum_{k}t_{\pm}^{k}\vec{\sigma}^{k} \parallel i> \mid ^{2}.
\end{equation}
Here $\vec{\sigma}^{k}$ is the spin operator and $t_{\pm}^{k}$
stands for the isospin raising and lowering operator and
$\left({g_{A} \mathord{\left/ {\vphantom {g_{A} g_{V} }} \right.
\kern-\nulldelimiterspace} g_{V} } \right)_{\emph{eff}}^{2}$ is the
effective ratio of the axial-vector $(g_{A} )$ to the vector $(g_{V}
)$ coupling constants that takes into account the observed quenching
of the GT strength (Osterfeld 1992). In the present work
$\left({g_{A} \mathord{\left/ {\vphantom {g_{A} g_{V} }} \right.
\kern-\nulldelimiterspace} g_{V} } \right)_{\emph{eff}}$ was taken
as
\begin{equation} \label{GrindEQ__4_}
\left({g_{A} \mathord{\left/ {\vphantom {g_{A}  g_{V} }} \right.
\kern-\nulldelimiterspace} g_{V} } \right)_{\emph{eff}} \, =\, 0.74
\left({g_{A} \mathord{\left/ {\vphantom {g_{A}  g_{V} }} \right.
\kern-\nulldelimiterspace} g_{V} } \right)_{\emph{bare}}
\end{equation}
with $\left({g_{A} \mathord{\left/ {\vphantom {g_{A}  g_{V} }}
\right. \kern-\nulldelimiterspace} g_{V} } \right)_{\emph{bare}}$
taken as -1.257. In the present work the dominant Gamow-Teller
transition strength are taken into account for the electron
capture/$\beta^{+}$-decay rate calculation.  In order to calculate
the Fermi transitions, the Coulomb displacement energy was
calculated as
\begin{equation}
\triangle E_{c} = (1.444 \acute{Z}/(A)^{1/3}) - 1.13 MeV.
\end{equation}
Where $\acute{Z}$ is average charge of the pair of the respective
nucleus and A is the mass number. The energy position of the
isobaric analog state (IAS) was computed using the following
equation:
\begin{equation}
E_{IAS} = \triangle E_{c} + \triangle - (m_{n} - m_{p})
\end{equation}
where $\triangle$ is the beta decay energy (to the ground state of
the daughter nucleus), $m_{n}$ and $m_{p}$ are mass of neutron and
proton, respectively.

The $f_{ij}^{ec(pd)}$ are the phase space integrals and are
functions of stellar temperature ($T$), electron density ($\rho
Y_{e}$) and Fermi energy ($E_{f}$) of the electrons. They are
explicitly given by
\begin{equation}
f_{ij}^{ec} \, =\, \int _{w_{l} }^{\infty }w\sqrt{w^{2} -1}
 (w_{m} \, +\, w)^{2} F(+Z,w)G_{-} dw.
 \label{ec}
\end{equation}
and by
\begin{equation}
f_{ij}^{pd} \, =\, \int _{1 }^{w_{m}}w\sqrt{w^{2} -1} (w_{m} \,
 -\, w)^{2} F(- Z,w)(1- G_{+}) dw,
 \label{pd}
\end{equation}
In Eqs. ~(\ref{ec}) and ~(\ref{pd}), $w$ is the total energy of the
electron including its rest mass. $w_{m}$ is the total $\beta$-decay
energy,
\begin{equation}
 w_{m} = \frac {1}{m_{e}c^{2}}(m_{p}-m_{d}+E_{i}-E_{j}),
\end{equation}
where $m_{p}$ and $E_{i}$ are masses and excitation energies of the
parent nucleus, and $m_{d}$ and $E_{j}$ of the daughter nucleus,
respectively. F($ \pm$ Z,w) are the Fermi functions and were
calculated according to the procedure adopted by Gove and Martin
(Gove et al. 1971). G$_{\pm}$ are the Fermi-Dirac distribution
functions for positrons (electrons).
\begin{equation}
G_{+} =\left[\exp \left(\frac{E+2+E_{f} }{kT}\right)+1\right]^{-1},
\end{equation}
\begin{equation}
 G_{-} =\left[\exp \left(\frac{E-E_{f} }{kT}
 \right)+1\right]^{-1},
\end{equation}
here $E$ is the kinetic energy of the electrons and $k$ is the
Boltzmann constant.

The neutrino blocking of the phase space was not taken into account
for reasons mentioned earlier. The total capture/$\beta^{+}$-decay
rate per unit time per nucleus is finally given by
\begin{equation}
\lambda^{ec(pd)} =\sum _{ij}P_{i} \lambda _{ij}^{ec(pd)}.
\label{total rate}
\end{equation}
The summation over the initial and final states was carried out
until satisfactory convergence was achieved is stellar weak rates
calculation. Here $P_{i} $ is the probability of occupation of
parent excited states and follows the normal Boltzmann distribution.

\section{Charge-changing Strength Distributions for Proton Drip-Line Nuclei}

Due to the advancement of Radioactive Ion Beam (RIB) facilities
worldwide (e.g. projectile- fragmentation facilities at GANIL with
powerful LISE3 separators for in-flight isotope separation), the
study of medium mass proton drip-line nuclei are now in experimental
reach (Pougheon et al. 1987; Borrel et al. 1992). As one moves on
the proton-rich side of the nuclear landscape the Q energy window in
the GT$_{+}$ (electron capture) direction increases and thereby
allows access to large GT strength. The pn-QRPA theory was used to
calculate the GT strength in the isospin raising direction, $T^{>}$,
for pf-shell nuclei at proton drip-line. The proton-rich nuclei
$^{46}$Mn, $^{48}$Fe, $^{50}$Co, $^{52}$Ni and $^{56}$Zn were
studied in this project and are also shown on the nuclear chart in
Fig.~1. These nuclei have an excess of four protons and represent
the bound nuclei (Pougheon et al. 1987).

The representative GT strength distributions in the electron capture
direction (GT$_{+}$) for $^{46}$Mn and $^{50}$Co are shown in Fig.~2
and Fig.~3, respectively. The energy scale refers to excitation
energy in daughter nuclei. The energy position of the isobaric
analogue state (IAS) is shown by dashed-line in Fig.~2 and Fig.~3.
The $E_{IAS}$ for $^{46}$Mn, $^{48}$Fe, $^{50}$Co, $^{52}$Ni, and
$^{56}$Zn is 25.06 MeV, 19.38 MeV, 25.75 MeV, 19.99 MeV, 22.09 MeV,
respectively. The ground state $Q_{ec}$ values and mass excess of
these nuclei with $T_{z} = -2$ were taken from the nuclear mass
compilations of Audi and Wapstra (Audi et al. 2003). Quenching
factor of 0.74 was used as employed in the shell model calculation
(Mart\'{i}nez-Pinedo et al. 1996; Langanke et al. 1999) for the
pf-shell nuclei. The $Q_{EC}$ value is represented by an arrow and
it is evident that most of the GT strength  structure appears within
the Q window of the reaction. A sharp decline in the GT strength can
be seen at high excitation energy beyond the Q window in the
daughter $^{46}$Cr (Fig.2). In this work a total unquenched and
quenched strength of 12.45 and 6.82, respectively, was calculated as
compared to the corresponding shell model values of 12.89 and 6.98,
respectively (Caurier et al. 1998).

For even-even $^{48}$Fe, the pn-QRPA extracted an unquenched and
quenched total GT strength of 12.62 and 6.91, respectively. This
quenched value is close to the quenched total strength of 7.18
reported by (Caurier et al. 1998). The mass-excess of the $^{48}$Fe
ground state, taken from (Audi et al. 2003), is taken to be (-18160
$\pm$ 70) keV.

For the odd-odd  nucleus $^{50}$Co the GT strength (shown in Fig.~3)
is fragmented over many states in the daughter  $^{50}$Fe due to the
correlations effect among the nucleons. The over all morphology of
the strength is in good agreement with calculated GT strength of
(Caurier et al. 1998). They used the KB3 interaction (Wang et al.
1988) to calculate the GT strength and these interactions are
modified version of Kuo-Brown interactions. The pn-QRPA extracted
more GT strength than calculated by shell model (Caurier et al.
1998). This decrease in their calculated GT strength could be due to
the truncated model space used in their calculation. It is noted
that bulk of the GT strength is distributed below or very close to
the Q window. The total GT strength is very crucial in the high
density regions in stellar core where the electrons are degenerate
and their Fermi energy grows faster than the corresponding Q value.
The pn-QRPA and shell model total unquenched and quenched GT
strengths for the proton drip-line nuclei are given in
Table~\ref{ta1}. These strengths were computed with the additional
assumption that the proton that has been converted to a neutron lie
within the same major shell.  It is noted that the pn-QRPA
calculated GT strength is in reasonable agreement with corresponding
shell model results. It is further noted that for heavier nuclei the
reported strengths are much bigger than shell model results (for
reasons mentioned above). The pn-QRPA calculated GT strength is
compared with experimental strength wherever available. Due to lack
of measurements of the ground state $\beta^{-}$ transition, one can
use the data for transitions of the mirror nucleus given by (p,n)
charge-exchange reactions. It is well known that the GT strength is
expected to be same in both isospin direction and the symmetry
equation $B(GT_{+}) = B(GT_{-})$ can be used for comparison. The
total GT strengths observed in the (p, n) charge-exchange reaction
(Wang et al. 1988; Rapaport et al. 1983) for the mirror nuclei
$^{52}$Cr and $^{56}$Fe with $T_{z} = +2$ are used for comparison in
Table~\ref{ta1}. It is seen that the pn-QRPA result is in good
agreement with the upper limit of the total strength observed in the
(p, n) charge-exchange reaction for the case of A = 56. At this
point, we reiterate the remarks in (Caurier et al. 1998) about the
reliability of the experimental data of these proton rich nuclei.
The strong splitting of the strength makes the observation more
difficult and thus imposes strict limits for the background in the
GT spectrum. The improvement in production rate of these proton rich
nuclei can assist in the possible observation of the shape of the
main components of the GT distribution by beta decay study.

The increase of GT$_{+}$ strength with mass number is natural as the
number of protons increases with the increase of mass number for the
selected proton drip-line nuclei. The total strength for these
nuclei (with $T_{z} = -2$ in the mass range $46\leq A \leq 56$) is
plotted as a function of mass number in Fig.~4. It is noted that
total strength increases roughly linearly with the increase of
proton number. This is attributed to the correlation effects between
the nucleons in-built within the QRPA model. The number of protons
increases as one moves from $^{46}$Mn to $^{56}$Zn along the proton
drip-line. This results in the enhancement of Coulombic repulsion
among the protons and consequently the Fermi surface of the proton's
orbital is pushed up and is close to the continuum.

\section{Electron Capture/$\beta^{+}$-Decay Rates for Proton Drip-Line Nuclei}
Bethe and coworkers (Bethe et al. 1979) pointed out the importance
of GT transitions for electron capture on heavy nuclei in the
presupernovae and collapse phases of stellar core. The entropy per
baryon in stellar core is very important as it determines the free
proton fraction. The proton-rich pf-shell nuclei may be used in the
studies of nucleosynthesis and energy generation in X-ray bursts and
other rp-process sites (Pruet et al. 2003; Wallace et al. 1981). The
authors in (Pruet et al. 2003) argued the need of weak rates on
proton-rich nuclei up to mass number 110 for nucleosynthesis in the
rp-process.

Electron capture rates based on independent particle model led to
incorrect conclusions due to the Pauli-blocking of the GT
transitions. This Pauli-blocking of the GT transitions is overcome
by the correlations effect among the nucleons (Caurier et al. 1999),
and temperature effects (Fuller et al. 1980; Cooperstein et al.
1984). This correlation effect is taken into account for calculation
of GT transitions in this project. The electron capture and
$\beta^{+}$ decay rates  for proton-rich pf-shell nuclei with $T_{z}
= - 2$ in the mass range 46 $\leq$ A $\leq$ 56, at various various
densities and temperatures, are given in Tables~\ref{ta2}-\ref{ta6}.
It has been noted that the $\beta^{+}$ decay rates dominate the
electron capture rates at low temperature and low densities in the
stellar interior. However, for densities $\geq 10^{9} gcm^{-3}$, the
electron capture process dominates even at low temperatures. Stellar
electron capture rates for proton-rich pf-shell nuclei $^{46}$Mn,
$^{48}$Fe, $^{50}$Co, $^{52}$Ni and $^{56}$Zn are shown in
Figs.~5-~9, respectively. These proton rich nuclei are characterized
by their large decay-values of 17.10 MeV, 11.16 MeV, 17.28 MeV,
11.27 MeV, and 12.87 MeV, respectively. On the other hand nuclei,
closer to the valley of stability, have usually smaller
decay-values. The electron capture rates for all five cases
(Figs.~5-~9) follow a more or less similar behavior which we explain
below. It is evident from discussion on GT strength in previous
section that a fair proportion of the GT strength lies at the upper
end of the Q-value window. This impedes the electron capture rates
in the low temperature and density regions of the stellar core. As
the temperature of the stellar matter increases in low density
regions the participant nucleon will get a fair chance of occupying
higher energy levels and in turn assists the electron captures on
the nucleon. For densities $\rho \leq 10^{11} g-cm^{-3}$, the
stellar weak rates are dominated by Fermi (applicable to beta decay
only) and GT transitions. There are two quantities that drive the
electron capture rates: the chemical potential of the electrons and
the nuclear Q-value. Chemical potential grows like $\rho^{1/3}$
(Nabi and Rahman 2005) and this growth of the chemical potential is
much faster than the Q-value of the nuclei in the stellar core.
Thus, the impedance posed by large Q-values of the reactions for
these nuclei is overcome by the fast growth of the chemical
potential of the electrons and consequently led to the enhancement
of the electron capture rates. At low densities, where the chemical
potential approximately equals nuclear Q-value, the capture rates
are very sensitive to the available phase space and detailed
description of the GT strength distribution is then desirable. It is
noted that in the low density region the beta decays compete with
the electron capture rates. These beta decays during and after
silicon shell burning increase the value of $Y_{e}$ in the stellar
core and cool the core efficiently as against competing electron
capture rates. The sensitivity to the phase space and details of the
GT strength distribution is less important as a result of fast
growth in chemical potential at high densities. The capture rates
are no more dependent on the details of GT strength distribution.
They rather depend on the total GT strength. In this scenario crude
nuclear models might also be able to give an estimate of the stellar
capture rates.

The experimentally measured and calculated half-lives for
proton-rich pf-shell nuclei with $T_{z} = - 2$ in the mass range 46
$\leq$ A $\leq$ 56 are mutually compared in Table~\ref{ta7}. The
pn-QRPA results are also compared with the results of the gross
theory of beta decay of (Tachibana et al. 1988) and (M\"{o}ller et
al. 1997). In addition our calculated half-lives are also compared
with the shell model calculation (Caurier et al. 1998). The results
of the pn-QRPA, for cases of $^{50}$Co and $^{56}$Zn, are in good
agreement with the shell model results and deviations are found for
the rest of the cases.

\section{Discussion and Conclusions}
Recent advancements in the accelerator driven technology has led to
noticeable improvement in nuclear inputs for core-collapse supernova
models. It has been noted that the collapse phase is dominated by
electron capture rates on nuclei rather than on free protons. The
GT$_{+}$ strength and associated electron capture/$\beta^{+}$-decay
rates were calculated within the domain of the pn-QRPA model. These
weak interaction mediated rates are key nuclear physics input to
simulation codes and a reliable and microscopic calculation of these
rates from ground-state and excited states is desirable. The
transitions in GT$_{+}$ direction and associated electron
capture/$\beta^{+}$-decay rates for pf-shell nuclei are important
from astrophysical point of view because these nuclei are key input
in the modeling of the explosion dynamics of massive stars. Our
calculation shows that $\beta^{+}$-decay rates dominate electron
capture rates at low stellar temperatures and densities.

The pn-QRPA theory with improved model parameters was used to
calculate weak-interaction mediated rates for the proton drip-line
nuclei. These calculations were carried out in a luxurious model
space of 7$\hbar\omega$. The pn-QRPA calculated half-lives of
$^{50}$Co and $^{56}$Zn are in reasonable agreement with shell model
calculation but differ in other cases. The GT strength distribution
is important for the description of the electron capture/$\beta^{+}$
rates related to pre-supernova and supernova conditions. The
pn-QRPA's total unquenched and quenched GT strength for proton
drip-line nuclei are compared with shell model values and with the
available experimental data. The pn-QRPA calculated GT strengths are
in reasonable agreement with the shell model and experimental data.
This may affect the evolution timescale and dynamics of collapsing
supermassive stars as the capture rates are dependent on the total
GT strength in the high temperature and density regions in stellar
interior. It has been found that total GT strength increases
linearly with increase of proton numbers as expected. The electron
capture/$\beta^{+}$-decay rates  for these proton drip-line nuclei,
for density and temperature scale relevant to astrophysical
scenario, may be requested from the authors as ASCII files.
Core-collapse simulators are encouraged to employ these rates in
simulation codes to check for possible interesting outcomes.

\onecolumn
\newpage

\begin{table}
\caption{Comparison of the pn-QRPA calculated GT$_{+}$ total
strength with shell model calculation (Caurier et al. 1998) and
measured GT strength for the pf-shell nuclei with $T_{z} = -2$ in
the mass region $46 \leq A \leq 56$.} \label{ta1}
\begin{tabular}
{c|cc|cc|c} Nucleus &  Unquenched
 & Unquenched  &
Quenched  & Quenched  & $\Sigma B(GT_{\pm})$ experimental\\
& $\Sigma B(GT_{+})^{pn-QRPA}$ & $\Sigma B(GT_{+})^{SM}$ & $ \Sigma
B(GT_{+})^{pn-QRPA}$ & $ \Sigma B(GT_{+})^{SM}$ & (mirror nuclei)
\\\\\hline
$^{46}$Mn   & 12.45 &    12.89 &   6.82 &   6.98 &   $-$\\
$^{48}$Fe   &  12.62&    13.26 &   6.91 &   7.18 &   $-$\\
$^{50}$Co   &  16.73&    14.68  &  9.16 &   7.95 &   $-$\\
$^{52}$Ni   &  17.51&    15.33 &   9.59 &   8.30 & 5.9 $\pm$ 1.5
(Wang 88)\\
$^{56}$Zn   & 21.91&    16.69 &   11.99&   9.04 & 9.9 $\pm$ 2.4
(Rapaport 83)\\
\end{tabular}
\end{table}

\begin{table}
\caption{$\beta^{+}$-decay rates and electron capture rates on
$^{46}$Mn for different selected densities and temperatures in
stellar matter. T$_{9}$ represents the temperature in $10^{9}$ K and
$\rho Y_{e}$ denotes the stellar density in units of $g/cm^{3}$.
Rates are given in log to base 10 scale and have units of $s^{-1}$.}
\label{ta2}
\begin{tabular}{cccc|cccc}
$\rho Y_{e} $ & $T_{9}$  & $\lambda_{\beta^{+}} $ & $\lambda_{ec}$ &
$\rho Y_{e} $ & $T_{9}$ &
$\lambda_{\beta^{+}} $ & $\lambda_{ec}$\\
\hline
10 & 0.01 & 0.852 & -5.000     & $10^{7}$ & 0.01 & 0.852 & -0.040\\
10 & 1 & 0.851 & -3.975      & $10^{7}$ & 1 & 0.851 & -0.044\\
10 & 3 & 0.915 & -1.411        & $10^{7}$ & 3 & 0.915 & 0.008\\
10 & 5 & 0.968 & -0.542         & $10^{7}$ & 5 & 0.969 & 0.090 \\
10 & 10 & 1.009 & 0.588         & $10^{7}$ & 10 & 1.011 & 0.681\\
10 & 30 & 0.977 & 2.536      & $10^{7}$ & 30 & 0.977 & 2.540\\
\hline
$10^{3}$ & 0.01 & 0.852 & -3.152  & $10^{9}$ & 0.01 & 0.852 & 2.213\\
$10^{3}$ & 1 & 0.851 & -3.748&      $10^{9}$ & 1 & 0.851 & 2.213\\
$10^{3}$ & 3 & 0.915 & -1.410     & $10^{9}$ & 3 & 0.915 &2.259\\
$10^{3}$ & 5 & 0.968 & -0.542     & $10^{9}$ & 5 & 0.970 &2.301\\
$10^{3}$ & 10 & 1.009 & 0.588     & $10^{9}$ & 10 & 1.019 &2.376\\
$10^{3}$ & 30 & 0.977 & 2.536     & $10^{9}$ & 30 & 1.012 &2.849\\
\hline
$10^{5}$ & 0.01 & 0.852 & -1.752  &$10^{11}$ & 0.01 & 0.852 & 5.040\\
$10^{5}$ & 1 & 0.851 & -1.942    & $10^{11}$ & 1 & 0.851 &5.032\\
$10^{5}$ & 3 & 0.915 & -1.353    & $10^{11}$ & 3 & 0.915 &5.083\\
$10^{5}$ & 5 & 0.968 & -0.533    & $10^{11}$ & 5 & 0.970 &5.121\\
$10^{5}$ & 10 & 1.009 & 0.589    & $10^{11}$ & 10 & 1.019 &5.156\\
$10^{5}$ & 30 & 0.977 & 2.536    & $10^{11}$ & 30 & 1.048 &
5.216\\\hline
\end{tabular}
\end{table}

\begin{table}
\caption{Same as Table~\ref{ta2} but for  $^{48}$Fe.} \label{ta3}
\begin{tabular}{cccc|cccc}
$\rho Y_{e} $ & $T_{9}$  & $\lambda_{\beta^{+}} $ & $\lambda_{ec}$ &
$\rho Y_{e} $ & $T_{9}$ & $\lambda_{\beta^{+}} $ & $\lambda_{ec}$\\
\hline
10 & 0.01 & -0.020 & -5.695     & $10^{7}$ & 0.01 & -0.020 & -0.689\\
10 & 1 & -0.020 & -4.656      & $10^{7}$ & 1 & -0.020 & -0.684\\
10 & 3 & -0.020 & -2.08        & $10^{7}$ & 3 & -0.020 & -0.637\\
10 & 5 & -0.002 & -1.175         & $10^{7}$ & 5 & -0.001 & -0.534 \\
10 & 10 & 0.714 & 0.232        & $10^{7}$ & 10 & 0.715 & 0.325\\
10 & 30 & 1.637 & 2.687      & $10^{7}$ & 30 & 1.638 & 2.691\\
\hline
$10^{3}$ & 0.01 & -0.020 & -3.847  & $10^{9}$ & 0.01 & -0.020 & 1.819\\
$10^{3}$ & 1 & -0.020 & -4.428&      $10^{9}$ & 1 & -0.020 & 1.826\\
$10^{3}$ & 3 & -0.020 & -2.079     & $10^{9}$ & 3 & -0.020 &1.836\\
$10^{3}$ & 5 & -0.002 & -1.175     & $10^{9}$ & 5 & 0.000 &1.857\\
$10^{3}$ & 10 & 0.714 & 0.232     & $10^{9}$ & 10 & 0.720 &2.065\\
$10^{3}$ & 30 & 1.637 & 2.688     & $10^{9}$ & 30 & 1.661 &2.997\\
\hline
$10^{5}$ & 0.01 & -0.020 & -2.445  &$10^{11}$ & 0.01 & -0.020 & 4.911\\
$10^{5}$ & 1 & -0.020 & -2.622    & $10^{11}$ & 1 & -0.020 &4.908\\
$10^{5}$ & 3 & -0.020 & -2.022    & $10^{11}$ & 3 & -0.020 &4.908\\
$10^{5}$ & 5 & -0.002 & -1.166    & $10^{11}$ & 5 & 0.000 &4.910\\
$10^{5}$ & 10 & 0.714 & 0.233    & $10^{11}$ & 10 & 0.720 &4.961\\
$10^{5}$ & 30 & 1.637 & 2.688    & $10^{11}$ & 30 & 1.685 & 5.302\\
\hline
\end{tabular}
\end{table}

\begin{table}
\caption{Same as Table~\ref{ta2} but for  $^{50}$Co.} \label{ta4}
\begin{tabular}{cccc|cccc}
$\rho Y_{e} $ & $T_{9}$  &
$\lambda_{\beta^{+}} $ & $\lambda_{ec}$ &
$\rho Y_{e} $ & $T_{9}$ & $\lambda_{\beta^{+}} $ & $\lambda_{ec}$\\
\hline
10 & 0.01 & 1.510 & -4.473     & $10^{7}$ & 0.01 & 1.510 & 0.462\\
10 & 1 & 1.544 & -3.427      & $10^{7}$ & 1 & 1.544 & 0.489\\
10 & 3 & 1.580 & -0.890        & $10^{7}$ & 3 & 1.580 & 0.523\\
10 & 5 & 1.594 & -0.055         & $10^{7}$ & 5 & 1.595 & 0.574 \\
10 & 10 & 1.600 & 1.020        & $10^{7}$ & 10 & 1.602 & 1.113\\
10 & 30 & 1.550 & 2.842      & $10^{7}$ & 30 & 1.550 & 2.846\\
\hline
$10^{3}$ & 0.01 & 1.510 & -2.625  & $10^{9}$ & 0.01 & 1.510 & 2.644\\
$10^{3}$ & 1 & 1.544 & -3.200     & $10^{9}$ & 1 & 1.544 & 2.681\\
$10^{3}$ & 3 & 1.544 & -0.889     & $10^{9}$ & 3 & 1.580 &2.709\\
$10^{3}$ & 5 & 1.594 & -0.054     & $10^{9}$ & 5 & 1.595 &2.726\\
$10^{3}$ & 10 & 1.600 & 1.020     & $10^{9}$ & 10 & 1.608 &2.769\\
$10^{3}$ & 30 & 1.550 & 2.842     & $10^{9}$ & 30 & 1.583 &3.151\\
\hline
$10^{5}$ & 0.01 & 1.510 & -1.228  &$10^{11}$ & 0.01 & 1.510 & 5.310\\
$10^{5}$ & 1 & 1.544 & -1.395    & $10^{11}$ & 1 & 1.544 &5.348\\
$10^{5}$ & 3 & 1.580 & -0.833    & $10^{11}$ & 3 & 1.580 &5.379\\
$10^{5}$ & 5 & 1.594 & -0.046    & $10^{11}$ & 5 & 1.595 &5.387\\
$10^{5}$ & 10 & 1.600 & 1.021    & $10^{11}$ & 10 & 1.608 &5.396\\
$10^{5}$ & 30 & 1.550 & 2.842    & $10^{11}$ & 30 & 1.617 & 5.435\\
\hline
\end{tabular}
\end{table}

\begin{table}
\caption{Same as Table~\ref{ta2} but for  $^{52}$Ni.} \label{ta5}
\begin{tabular}{cccc|cccc}
$\rho Y_{e} $ & $T_{9}$  & $\lambda_{\beta^{+}} $ & $\lambda_{ec}$ &
$\rho Y_{e} $ & $T_{9}$ & $\lambda_{\beta^{+}} $ & $\lambda_{ec}$\\
\hline
10 & 0.01 & 0.810 & -5.162     & $10^{7}$ & 0.01 & 0.810 & -0.207\\
10 & 1 & 0.810 & -4.141      & $10^{7}$ & 1 & 0.810 & -0.206\\
10 & 3 & 0.809 & -1.604        & $10^{7}$ & 3 & 0.810 & -0.178\\
10 & 5 & 0.813 & -0.740         & $10^{7}$ & 5 & 0.814 & -0.105 \\
10 & 10 & 0.972 & 0.507        & $10^{7}$ & 10 & 0.973 & 0.601\\
10 & 30 & 1.549 & 2.866      & $10^{7}$ & 30 & 1.549 & 2.870\\
\hline
$10^{3}$ & 0.01 & 0.810 & -3.314  & $10^{9}$ & 0.01 & 0.810 & 2.155\\
$10^{3}$ & 1 & 0.810 & -3.913     & $10^{9}$ & 1 & 0.810 & 2.161\\
$10^{3}$ & 3 & 0.809 & -1.603     & $10^{9}$ & 3 & 0.810 &2.169\\
$10^{3}$ & 5 & 0.813 & -0.740     & $10^{9}$ & 5 & 0.814 &2.186\\
$10^{3}$ & 10 & 0.972 & 0.508     & $10^{9}$ & 10 & 0.978 &2.327\\
$10^{3}$ & 30 & 1.549 & 1.549     & $10^{9}$ & 30 & 1.574 &3.178\\
\hline
$10^{5}$ & 0.01 & 0.810 & -1.917  &$10^{11}$ & 0.01 & 0.810 & 5.120\\
$10^{5}$ & 1 & 0.810 & -2.108    & $10^{11}$ & 1 & 0.810 &5.116\\
$10^{5}$ & 3 & 0.809 & -1.546    & $10^{11}$ & 3 & 0.810 &5.117\\
$10^{5}$ & 5 & 0.813 & -0.731    & $10^{11}$ & 5 & 0.814 &5.119\\
$10^{5}$ & 10 & 0.972 & 0.508    & $10^{11}$ & 10 & 0.978 &5.170\\
$10^{5}$ & 30 & 1.549 & 2.866    & $10^{11}$ & 30 & 1.598 &
5.517\\\hline
\end{tabular}
\end{table}

\begin{table}
\caption{Same as Table~\ref{ta2} but for  $^{56}$Zn.} \label{ta6}
\begin{tabular}{cccc|cccc}
$\rho Y_{e} $ & $T_{9}$  &
$\lambda_{\beta^{+}} $ & $\lambda_{ec}$ &
$\rho Y_{e} $ & $T_{9}$ & $\lambda_{\beta^{+}} $ & $\lambda_{ec}$\\
\hline
10 & 0.01 & 1.399 & -4.429     & $10^{7}$ & 0.01 & 1.399 & 0.489\\
10 & 1 & 1.399 & -3.421      & $10^{7}$ & 1 & 1.399 & 0.488\\
10 & 3 & 1.399 & -0.913        & $10^{7}$ & 3 & 1.399 & 0.499\\
10 & 5 & 1.398 & -0.085         & $10^{7}$ & 5 & 1.399 & 0.544 \\
10 & 10 & 1.390 & 0.989        & $10^{7}$ & 10 & 1.392 & 1.082\\
10 & 30 & 1.331 & 2.822      & $10^{7}$ & 30 & 1.332 & 2.826\\
\hline
$10^{3}$ & 0.01 & 1.399 & -2.581  & $10^{9}$ & 0.01 & 1.399 & 2.685\\
$10^{3}$ & 1 & 1.399 & -3.194     & $10^{9}$ & 1 & 1.399 & 2.691\\
$10^{3}$ & 3 & 1.399 & -0.912     & $10^{9}$ & 3 & 1.399 &2.695\\
$10^{3}$ & 5 & 1.398 & -0.085     & $10^{9}$ & 5 & 1.399 &2.705\\
$10^{3}$ & 10 & 1.390 & 0.990     & $10^{9}$ & 10 & 1.399 &2.745\\
$10^{3}$ & 30 & 1.331 & 2.822     & $10^{9}$ & 30 & 1.365 &3.132\\
\hline
$10^{5}$ & 0.01 & 1.399 & -1.188  &$10^{11}$ & 0.01 & 1.399 & 5.379\\
$10^{5}$ & 1 & 1.399 & -1.389    & $10^{11}$ & 1 & 1.399 &5.376\\
$10^{5}$ & 3 & 1.399 & -0.855    & $10^{11}$ & 3 & 1.399 &5.376\\
$10^{5}$ & 5 & 1.398 & -0.076    & $10^{11}$ & 5 & 1.399 &5.377\\
$10^{5}$ & 10 & 1.390 & 0.991    & $10^{11}$ & 10 & 1.399 &5.381\\
$10^{5}$ & 30 & 1.331 & 2.822    & $10^{11}$ & 30 & 1.399 &
5.417\\\hline
\end{tabular}
\end{table}

\begin{table}
\caption{Comparison of half-lives of the pn-QRPA model with
experimental: (a) Faux et al. 1994, (b) Borrel et al. 1992, (c) Faux
et al. 1996, (d) Dossat et al. 2007, (e) Audi et al. 2003b and other
theoretical calculations : (f) Caurier et al. 1998, (g) Tachibana
et. al 1988 and (h) M\"{o}ller et.al 1997 for the pf-shell nuclei
with $T_{z} = -2$ in the mass region $46 \leq A \leq 56$. All
half-lives are given in units of $s^{-1}$.} \label{ta7}
\begin{tabular}{cccccccc}
Nucleus & [T$_{1/2}]^{Exp.}$
 & [T$_{1/2}]^{(e)}$ & [T$_{1/2}]^{pn-QRPA}$ & [T$_{1/2}]^{(f)}$
 & [T$_{1/2}]^{(g)}$ & [T$_{1/2}]^{(h)}$
\\\hline
$^{46}$Mn   & $ 41 ^{+7{(a)}}_{-6}$ &    37 &   97.4 &  29 & 53 & 15 \\
$^{48}$Fe   &  $44\pm 7^{(b)}$ &   44 &   125.8 &   53 &    48 & 60 \\
$^{50}$Co   &  $44\pm4^{(b)}$ &    44  &  21.4 &  27 &   36  & 47 \\
$^{52}$Ni   &  $38\pm 5^{(c)}$&    38 &   107.3 &   50 & 35   & 77  \\
& $40.08\pm 2^{(d)}$ & & & & &  \\
$^{56}$Zn   & $-$ &    36 &   27.6 &   24 & 24 & 83 \\
\end{tabular}
\end{table}
 \newpage
\begin{figure}
\begin{center}
\includegraphics[width=6in,height=5.5in]{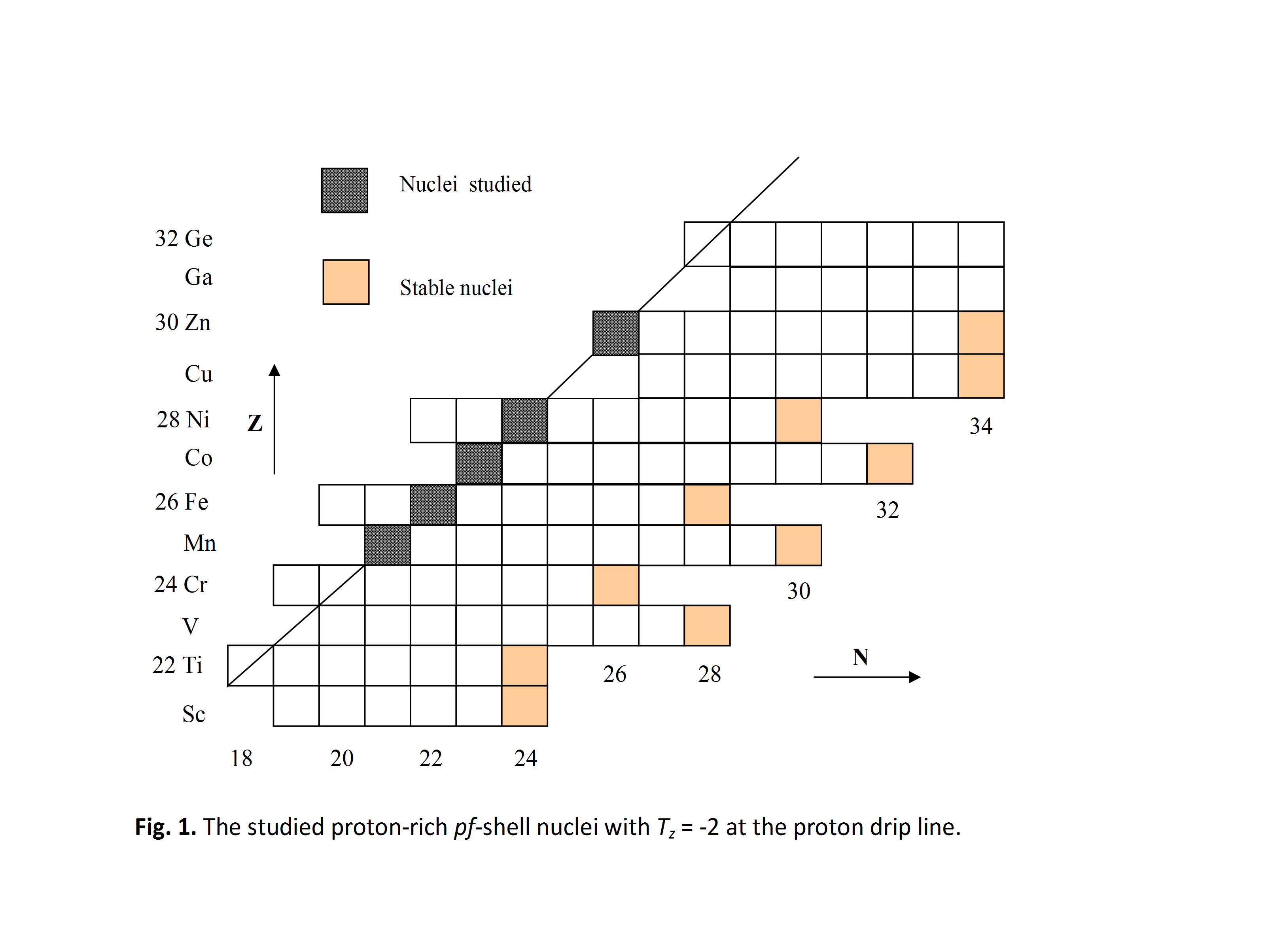}
\caption{(Color online) The studied proton-rich pf-shell nuclei with
$T_{z} = -2$ at the proton drip-line.}\label{fig1}
\end{center}
\end{figure}
\begin{figure}
\begin{center}
\includegraphics[width=6in,height=5.5in]{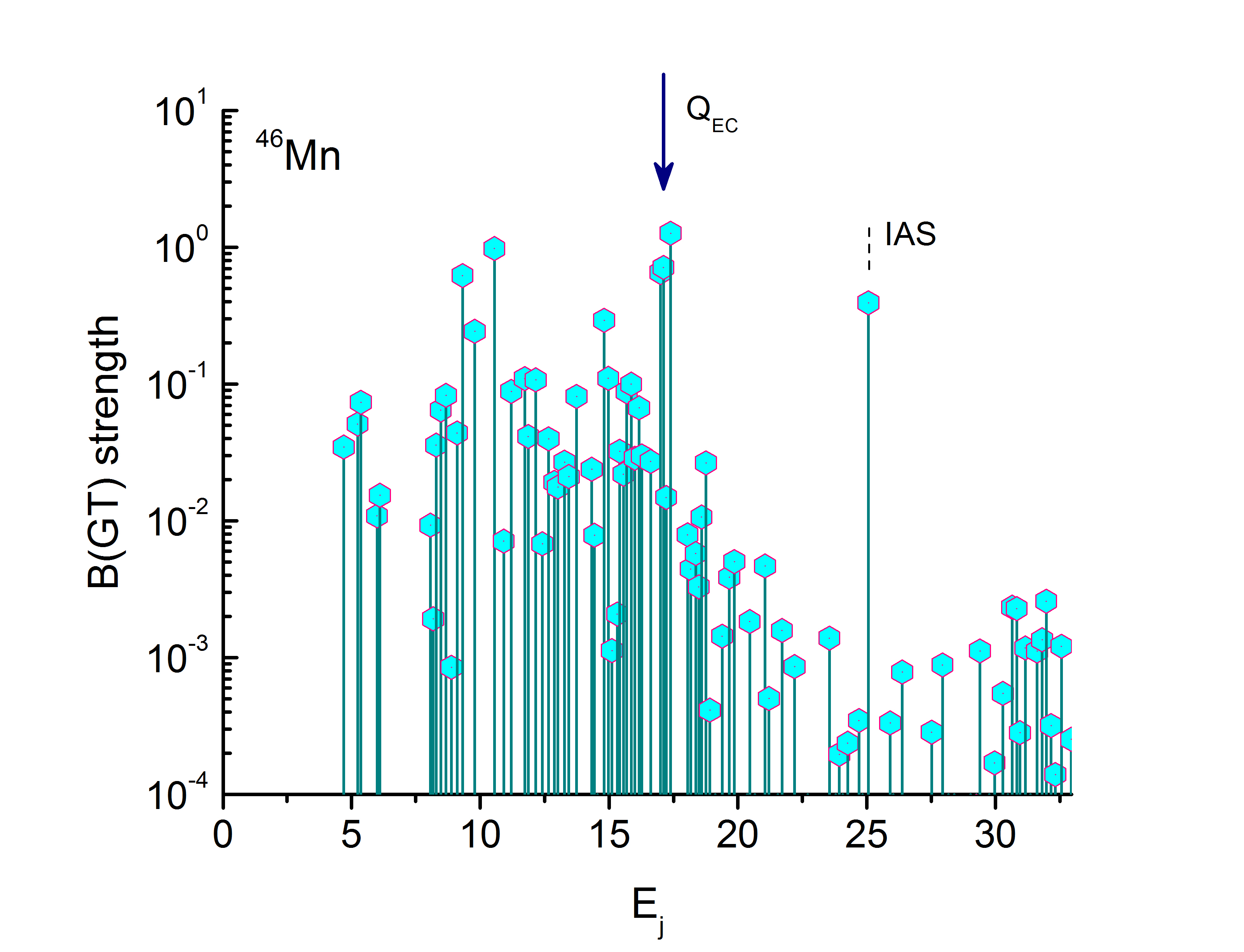}
\caption{(Color online) Calculated GT strength distribution for
$^{46}$Mn. E$_{j}$ refers to the excitation energy (MeV) in daughter
$^{46}$Cr. The isobaric analog state (IAS) is shown with a dashed
line while the Q-value for the electron capture reaction is shown by
solid arrow.}\label{fig2}
\end{center}
\end{figure}
\begin{figure}
\begin{center}
\includegraphics[width=6.0in,height=5.5in]{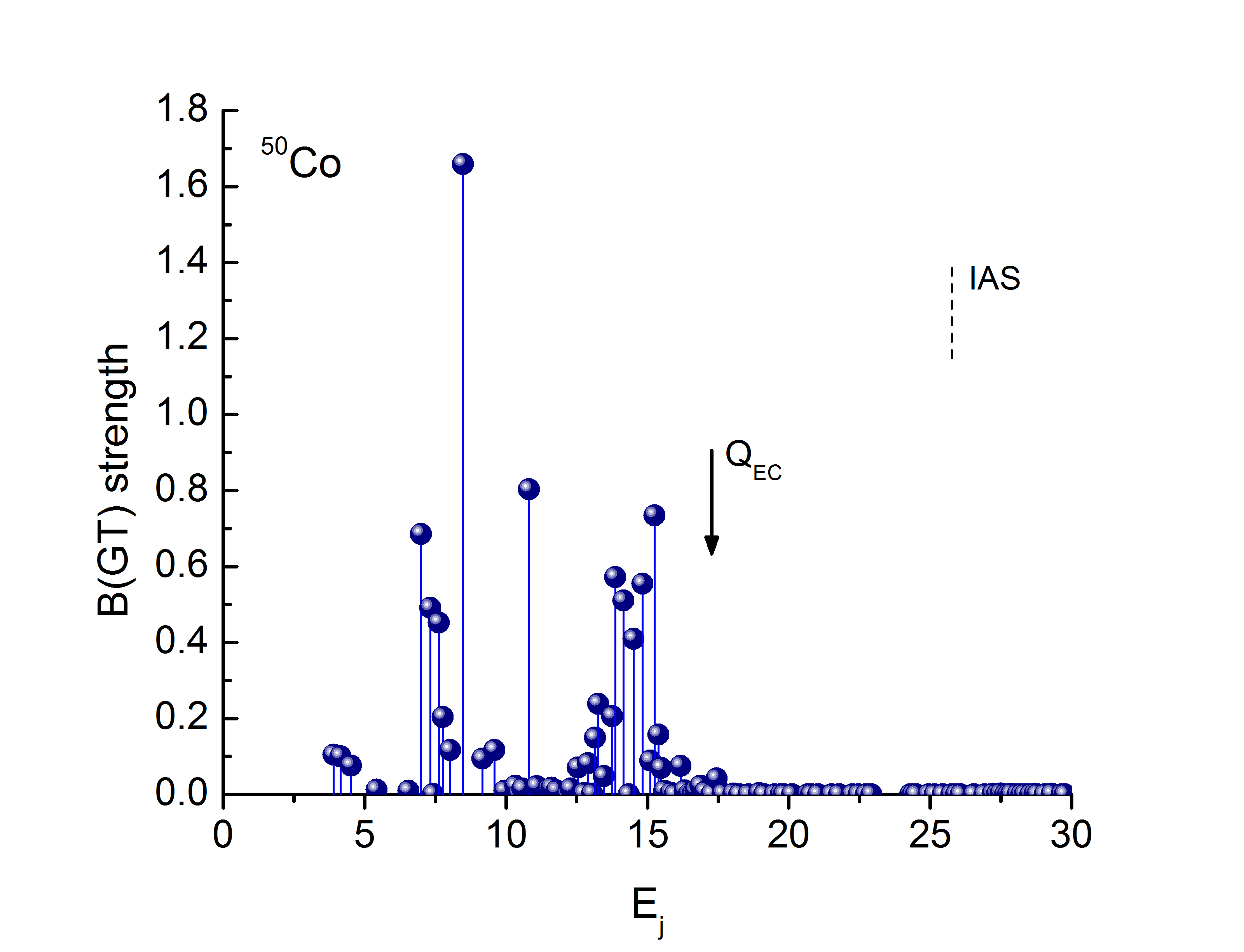} \caption { (Color online) Calculated GT strength distribution for $^{50}$Co. The
energy scale refers to the excitation energy (MeV) in daughter
$^{50}$Fe. The isobaric analog state (IAS) is shown with a dashed
line while the Q-value for the electron capture reaction is shown by
solid arrow.}\label{fig3}
\end{center}
\end{figure}
\begin{figure}
\begin{center}
\includegraphics[width=6in,height=5.5in]{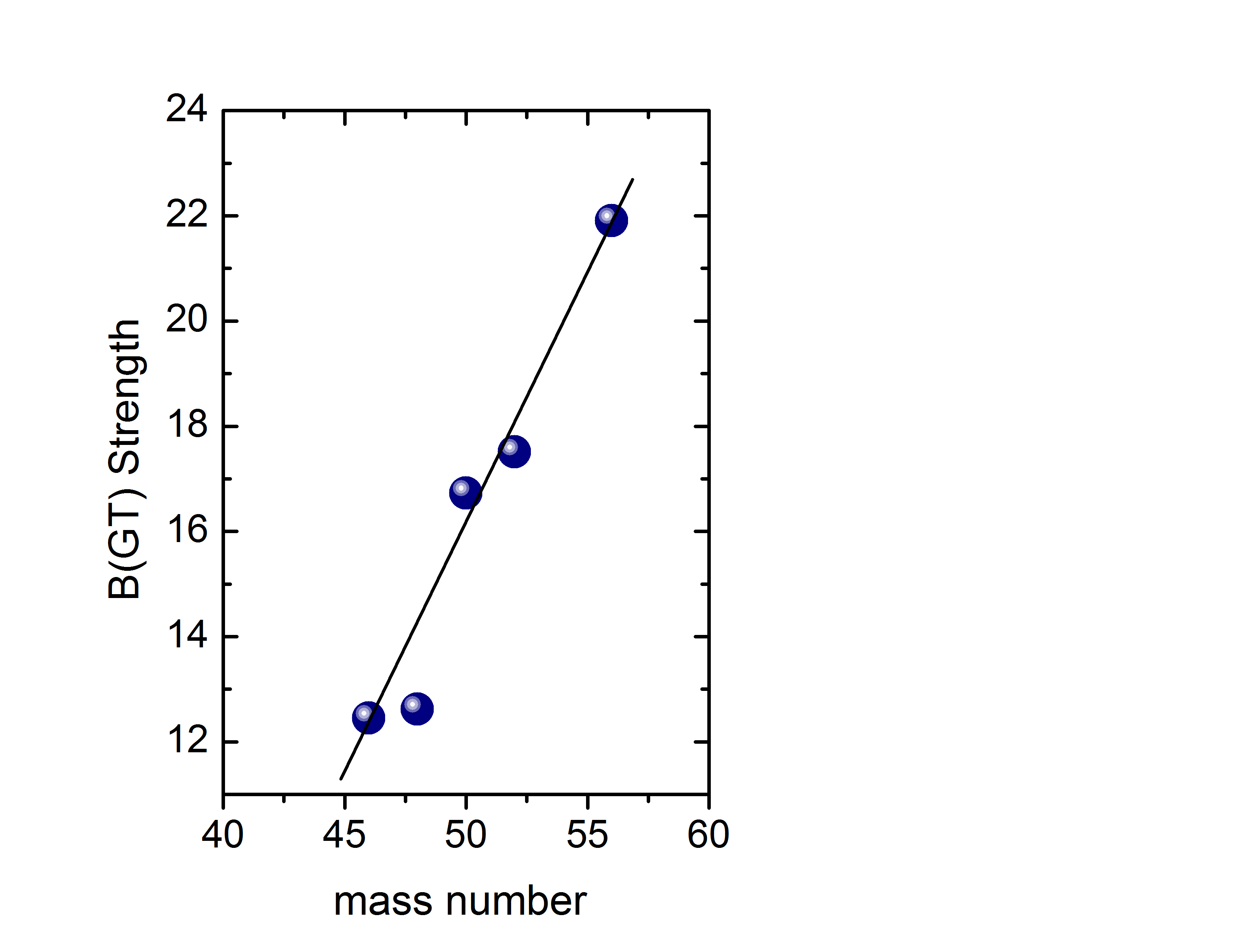} \caption {(Color online) Calculated total GT strength as a function of mass
number for the $T_{z} = -2$ nuclei in the mass region $46 \leq A
\leq 56$.}\label{fig4}
\end{center}
\end{figure}
\begin{figure}
\begin{center}
\includegraphics[width=6in,height=5.5in]{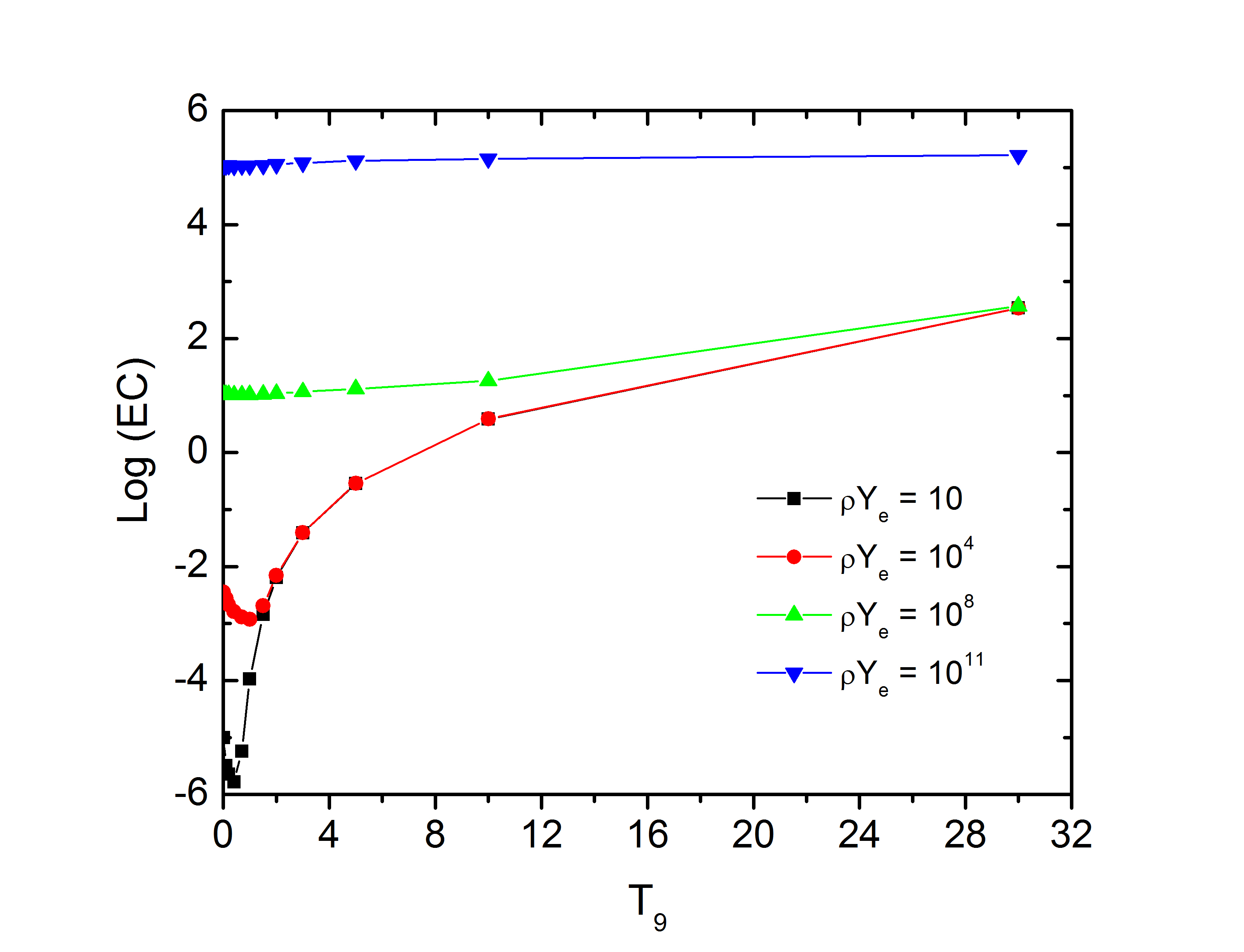} \caption {(Color online) Stellar electron capture rates for $^{46}$Mn
as a function of temperature. Electron capture (EC) rates are given
in log (to base 10) scale in units of $s^{-1}$. T$_{9}$ represents
temperature in units of $10^{9}$ K. Densities in inset are given in
units of g-cm$^{-3}$.}\label{fig5}
\end{center}
\end{figure}
\begin{figure}
\begin{center}
\includegraphics[width=6in,height=5.5in]{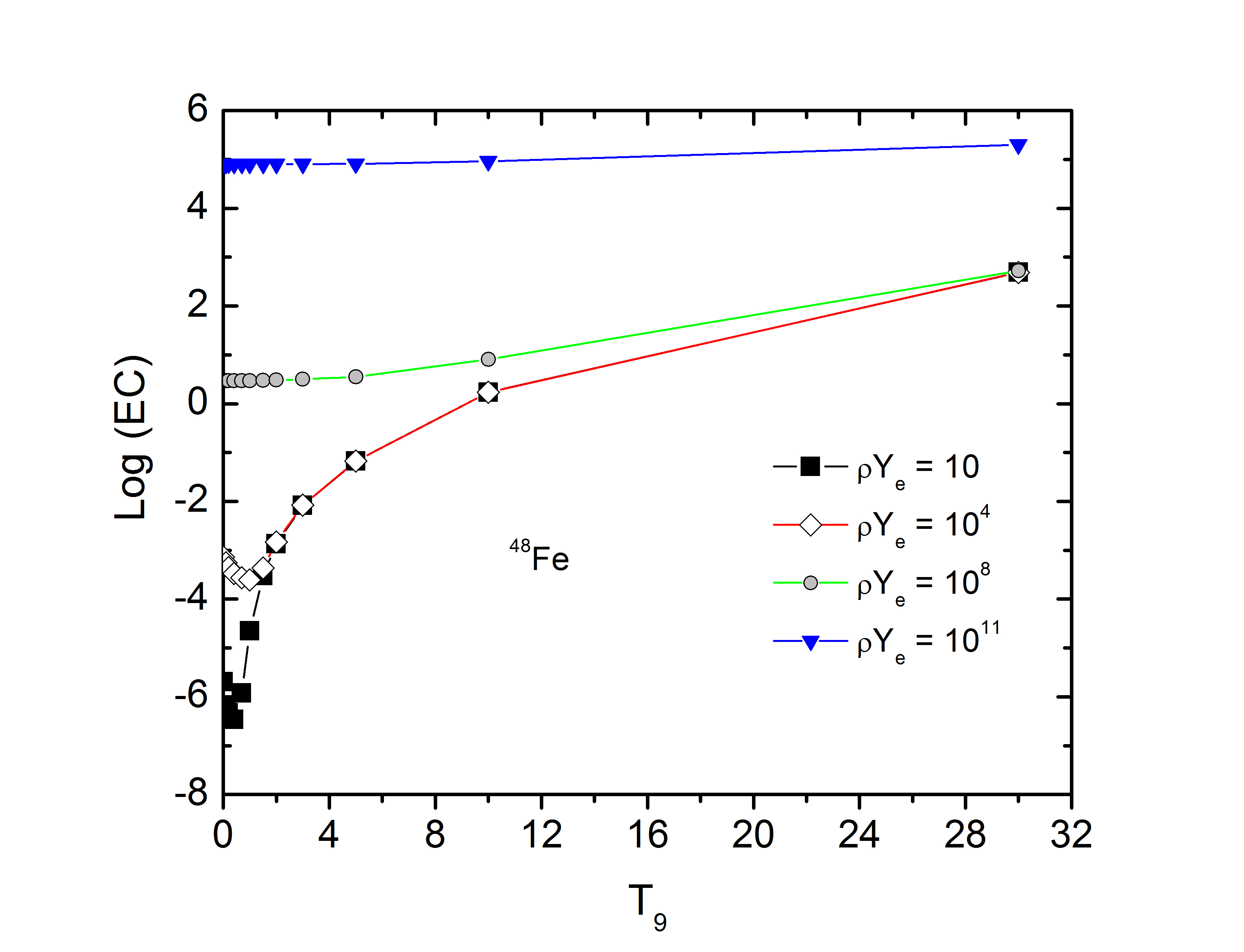} \caption {(Color online) Same as Fig.~5 but for $^{48}$Fe.}\label{fig6}
\end{center}
\end{figure}
\begin{figure}
\begin{center}
\includegraphics[width=6in,height=5.5in]{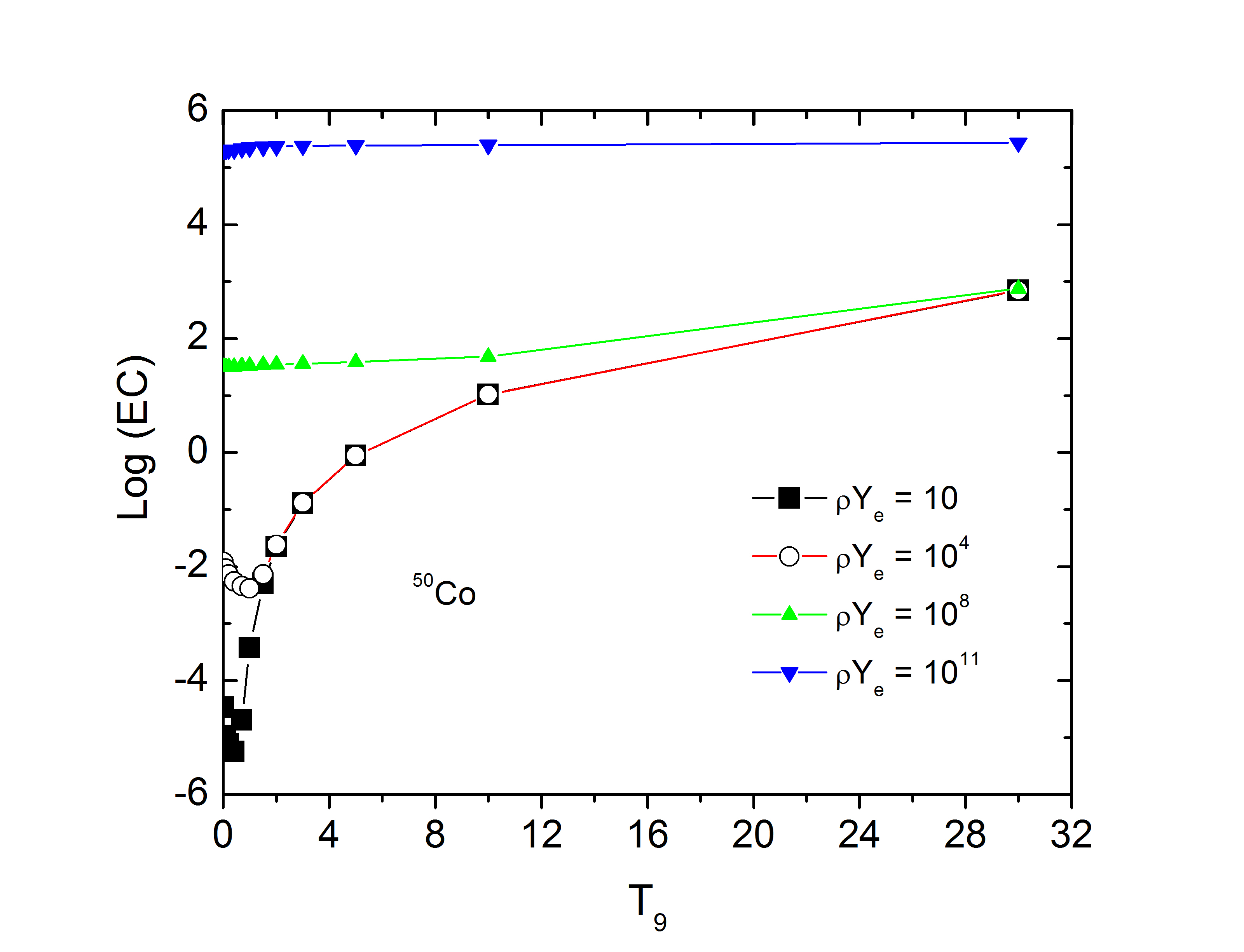} \caption {(Color online) Same as Fig.~5 but for $^{50}$Co.}\label{fig7}
\end{center}
\end{figure}
\begin{figure}
\begin{center}
\includegraphics[width=6in,height=5.5in]{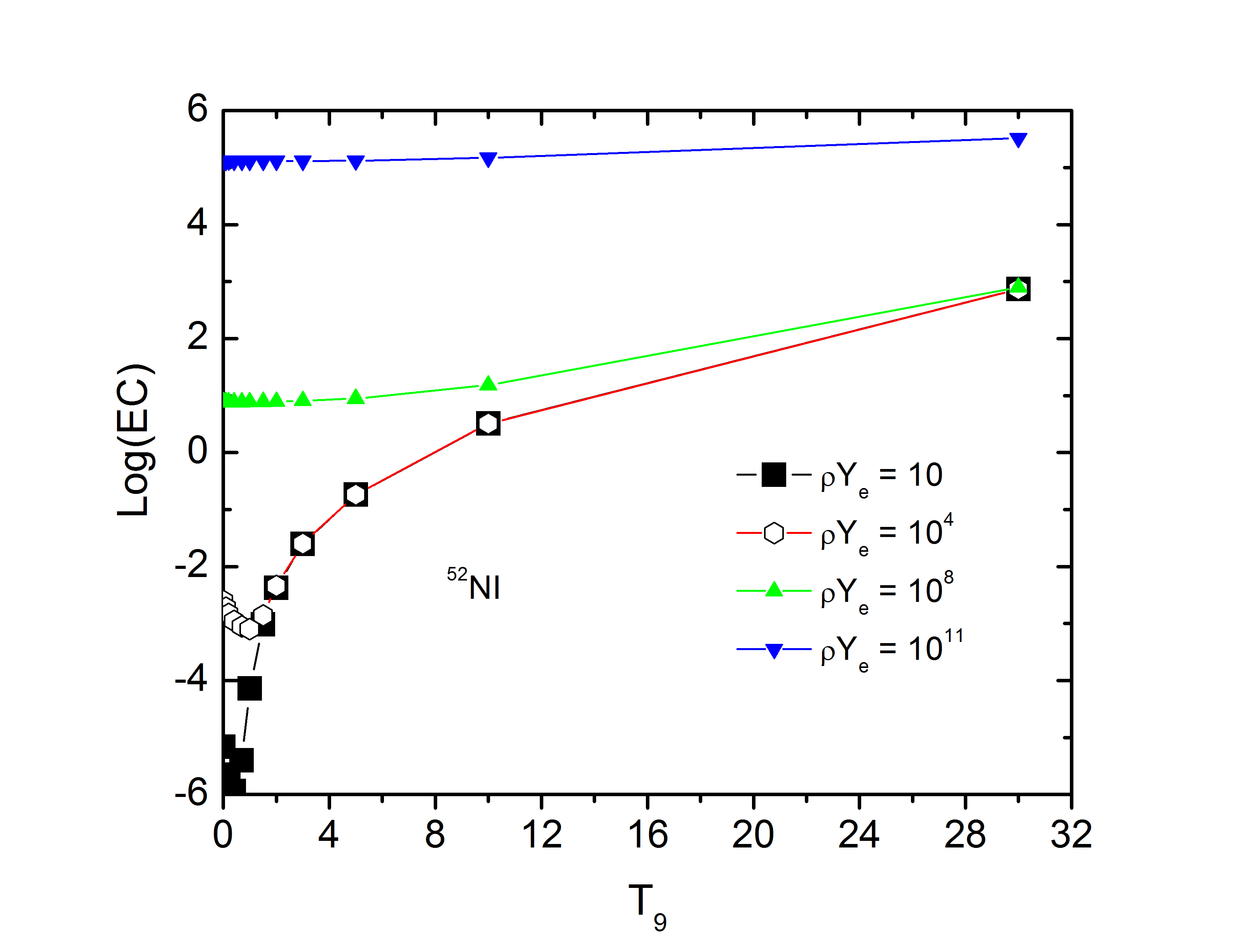} \caption {(Color online) Same as Fig.~5 but for $^{52}$Ni.}\label{fig8}
\end{center}
\end{figure}
\begin{figure}
\begin{center}
\includegraphics[width=6in,height=5.5in]{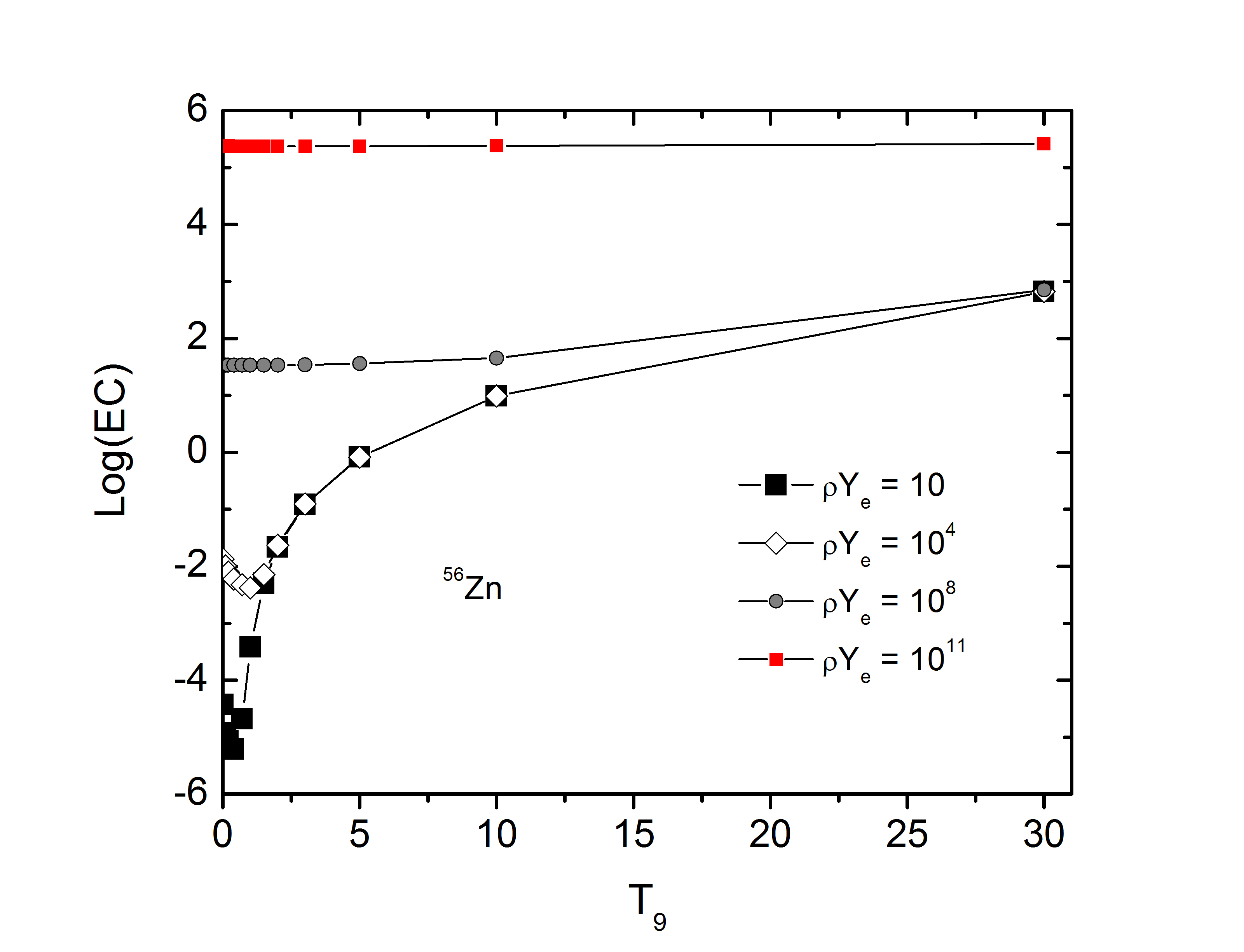} \caption {(Color online) Same as Fig.~5 but for $^{56}$Zn.}\label{fig9}
\end{center}
\end{figure}


\begin{thebibliography}{99}
\bibitem[Anderson et al. 1990]{Anderson90} Anderson, B.D., Lebo, C., Baldwin, A.R.,
Chittrakarn, T., Mandey, R., Watson, J.W. 1990, Phys. Rev. C,
\textbf{41}, 1474
\bibitem[Aufderheide et al. 1996]{Aufd96} Aufderheide, M.B., Bloom, S.D., Mathews, G.J.,
Resler, D.A. 1996, Phys. Rev. C, \textbf{53(6)}, 3139-3142
\bibitem[Audi et al. 2003a]{Audi03a} Audi, G., Wapstra, A.H., Thibault, C. 2003, Nucl.
Phys. A, \textbf{729}, 337
\bibitem[Audi et al. 2003b]{Audi03b} Audi, G., Bersillon, O., Blachot, J., Wapstra, A.H. 2003, Nucl.
Phys. A, \textbf{729}, 3-128
\bibitem[Bethe et al. 1979]{Bethe79}Bethe, H.A., Brown, G.E., Applegate, J., Lattimer,
J. 1979, Nucl. Phys. A, \textbf{324}, 487
\bibitem[Bethe 1990]{Bethe90} Bethe, H.A. 1990, Rev. Mod. Phys., \textbf{62}, No. 4, 801-866
\bibitem[Borrel et al. 1992]{Borrel992} Borrel, V.,  Anne, R., Bazin, D., Borcea, C.
Chubarian, G. G., Del Moral, R., Détraz, C., Dogny, S., Dufour,
J.P., Faux, L., Fleury, A., Fifield, L.K., Guillemaud-Mueller, D.,
Hubert, F., Kashy, E., Lewitowicz, M., Marchand, C., Mueller, A.C.,
Pougheon, F., Pravikoff, M.S., Saint-Laurent, M.G., Sorlin, O. 1992,
Z. Phys. A - Hadrons and Nuclei, \textbf{344}, 135
\bibitem[cole et al. 2012]{cole12}Cole, A.L., Anderson, T.S., Zegers, R.G.T., Austin, Sam M., Brown, B.A.,
Valdez, L., Gupta, S., Hitt, G.W., Fawwaz, O. 2012, Phys. Rev. C,
\textbf{86}, 015809
\bibitem[Cooperstein et al. 1984]{Cooperstein84}Cooperstein, J., Wambach, J. 1984, Nucl. Phys.
A, \textbf{420}, 591-620
\bibitem[Caurier et al 1998]{Caurier98} Caurier, E., Didierjean, F., Nowacki, F., Walter, G. 1998, Phys. Rev. C, \textbf{57}, 2316
\bibitem[Caurier et al. 1999]{Caurier99}Caurier, E., Langanke, K., Mart\'{i}nez-Pinedo,
G., Nowacki, F. 1999, Nucl. Phys. A, \textbf{653}, 439-452
\bibitem[Dossat et al. 2007]{Dossat07} Dossat, C., Blank, B.  et al. 2007, Nucl. Phys. A, \textbf{792}, 18
\bibitem[El-Kateb et al. 1994]{El-Kateb94} El-Kateb, S., Jackson, K.P., Alford, W.P., Abegg, R., Azuma, R.E., Brown, B.A., Celler, A.,
Frekers, D., Häusser, O., Helmer, R., Henderson, R.S., Hicks, K.H.,
Jeppesen, R., King, J.D., Shute, G.G., Spicer, B.M., Trudel, A.,
Raywood, K., Vetterli, M., Yen, S. 1994, Phys. Rev. C, \textbf{49},
3129
\bibitem[Faux et al. 1994] {Faux94} Faux, L., et al., 1994, Phys. Rev. C \textbf{49}, 2440
\bibitem[Faux et al. 1996] {Faux96} Faux, L., et al., 1996, Nucl. Phys. A \textbf{602}, 167
\bibitem[Fuller et al. 1980]{FFN80} Fuller, G.M., Fowler, W.A., Newman, M.J. 1980,
\apjs, \textbf{42}, 447 (1980).
\bibitem[Fuller et al. 1982a]{FFN82a} Fuller, G.M., Fowler, W.A., Newman, M.J. 1982a, \apjs, \textbf{48}
279
\bibitem[Fuller et al. 1982b]{FFN82b} Fuller, G.M., Fowler, W.A., Newman, M.J. 1982b, \apj, \textbf{252}, 715
\bibitem[Fuller et al.1985]{FFN85} Fuller, G.M., Fowler, W.A., Newman, M.J. 1985, \apj, \textbf{293}, 1-16
\bibitem[Gove et al. 1971]{Gove71} Gove, N.B., Martin, M.J. 1971, Nucl. Data Tables, \textbf{10}, 205
\bibitem[Hirsch et al. 1993]{Hirsch93} Hirsch, M., Staudt, A., Muto, K.,
Klapdor-Kleingrothaus, H.V. 1993, At. Data Nucl. Data Tables,
\textbf{53}, 165
\bibitem[Heger et al. 2001]{Heger01} Heger, A., Woosley, S.E., Mart\'{i}nez-Pinedo,
G., Langanke, K. 2001, The Astrophysical Journal, \textbf{560},
307-325
\bibitem[Jokinen et al. 2002]{Jokinen02} Jokinen, A., Nieminen, A., Äystö, J., Borcea, R.,
Caurier, E., Dendooven, P., Gierlik, M., Górska, M., Grawe, H.,
Hellström, M., Karny, M., Janas, Z., Kirchner, R., La Commara, M.,
Mart\'{i}nez-Pinedo, G., Mayet, P., Penttilä, H., Plochocki, A.,
Rejmund, M., Roeckl, E., Sawicka, M., Schlegel, C., Schmidt, K.,
Schwengner, R. 2002, Euro. Phys. J. Direct A \textbf{3}, 1
\bibitem[Juodagalvis et al. 2010]{Juodagalvis10}Juodagalvis, A., Langanke, K., Hix, W.R.,
Mart\'{i}nez-Pinedo, G., Sampaio, J.M. 2010, Nucl. Phys. A,
\textbf{848}, 454-478
\bibitem[Liu 2013]{Liu13} Jin-Jing Liu 2013, \mnras, \textbf{433}, 1108-1113
\bibitem[Kar et al. 1994]{Kar94}Kar, K., Ray, A., Sarkar, S. 1994, The Astrophysical Journal, \textbf{434}, 662-683
\bibitem[Langanke et al. 1998]{Langanke98} Langanke, K., Mart\'{i}nez-Pinedo,
G. 1998, Phys. Lett. B, \textbf{436}, 19
\bibitem[Langanke 1999]{Langanke99} Langanke, K., Mart\'{i}nez-Pinedo, G. 1999, Phys. Lett. B, \textbf{453}, 187
\bibitem[M\"{o}ller et al. 1981]{Moe81}M\"{o}ller P., Nix J.R. 1981, At. Data Nucl. Data
Tables, \textbf{26}, 165
\bibitem[M\"{o}ller et al. 1997]{Moe97} M\"{o}ller, P., Nix, J.R., Kratz, K.-L. 1997,
At. Nucl. Data Tables, \textbf{66}, 131
\bibitem[Muto et al. 1992]{Muto92}  Muto, K., Bender, E., Oda, T., Klapdor-Kleingrothaus, H. V. 1992, Z. Phys.
A, \textbf{341}, 407
\bibitem[Mart\'{i}nez-Pinedo et al. 1996]{Mar96} Mart\'{i}nez-Pinedo, G., Poves, A., Caurier, E.,
Zuker, A.P. 1996, Phys. Rev. C, \textbf{53}, 2602-2605
\bibitem[Nilsson 1955]{Nilsson55} Nilsson, S.G. 1955, Mat. Fys. Medd.
Dan. Vid. Selsk, \textbf{29}, 16
\bibitem[Nabi et al. 1999a]{Nab99a} Nabi J.-Un, Klapdor-Kleingrothaus H.V. 1999a, Eur. Phys. J. A, \textbf{5}, 337
\bibitem[Nabi et al. 2004]{Nab04} Nabi J.-Un, Klapdor-Kleingrothaus H.V. 2004, At. Data Nucl. Data Tables, \textbf{ 88},
237
\bibitem[Nabi and Rahman 2005]{Nabi and Rahman05} Nabi, J.-U., Rahman, M.-U. 2005, Phys. Lett. B, \textbf{612}, 190-196
\bibitem[Nabi 2009]{Nab09}Nabi J.-Un. 2009, Eur.
Phys. J. A, \textbf{40}, 223
\bibitem[Nabi 2012]{Nab12}Nabi J.-Un. 2012, Astrophys. and Space Sci., \textbf{339},
305-315
\bibitem[Nabi et al. 2013]{Nab13}Nabi J.-Un. and Johnson C.W. 2013, J. Phys. G, \textbf{40},
065202
\bibitem[Osterfeld 1992]{Osterfeld92} Osterfeld, F. 1992, Rev. Mod. Phys. \textbf{64}, 491
\bibitem[Pougheon et al. 1987]{Pougheon87} Pougheon, F., Jacmart, J.C., Quiniou, E., Anne, R., Bazin, D., Borrel, V.,
Galin, J., Guerreau, D., Guillemaud-MueUer, D., Mneller, A.C.,
Roeckl, E., Saint-Laurent, M.G., Détraz, C. 1987, Z. Phys. A -
Atomic Nuclei, \textbf{327}, 17
\bibitem[Pruet et al. 2003]{Pruet03} Pruet, J., Fuller, G.M. 2003, Astrophys. J. Suppl. \textbf{149}, 189
\bibitem[Rapaport et al. 1983]{Rapaport83}Rapaport, J., Taddeucci, T., Welch, T.P., Gaarde, C., Larsen, J., Horen, D.J.,
Sugarbaker, E., Koncz, P., Foster, C.C., Goodman, C.D., Goulding,
C.A., Masterson, T. 1983, Nucl. Phys. A, \textbf{410}, 371
\bibitem[Ronnqvist et al. 1993]{Ronnqvist93} Ronnquist, T., Cond\'{e}, H., Olsson, N., Remstr$\odot{o}$m, E., Zorro, R.,
Blomgreen, J., H$\dot{a}$kansson, A., Ringbom, A., Tibell, G.,
Jonsson, O., Nilsson, L., Renberg, P.–U., Van der Werf, S.Y.,
Unkelbach, W., Brady, F.P. 1993, Nucl. Phys. A, \textbf{563}, 225
\bibitem[Rahman et al. 2013]{Rahman13}Rahman, M.-U, Nabi J.-U. 2013, Astrophys. and Space
Sci., \textbf{348}, 427-435
\bibitem[Rahman et al. 2014]{Rahman14}Rahman, M.-U, Nabi, J.-U. 2014, Astrophys. Space Sci., \textbf{351}, 235-242,
DOI:10.1007/s10509-014-1831-0.
\bibitem[Sarriguren 2013]{Sarriguren13}Sarriguren, P., 2013, Phys. Rev. C, \textbf{87}, 045801
\bibitem[Staudt et al. 1990]{Staudt90} Staudt, A., Hirsch, M., Muto, K., Klapdor-Kleingrothaus, H. V. 1990, Phys.
Rev. Lett., \textbf{65}, 1543
\bibitem[Tachibana et al. 1988]{Tachibana88} Tachibana, T., Yamada, M., Nakata, K., Report of
Sci. and Res. Lab., Waseda University, No. 88-3 (1988)
\bibitem[Timmes et al. 1996]{Timmes96}Timmes, F.X., Woosley, S.E., Weaver, T.A. 1996,
\apj, \textbf{457}, 834
\bibitem[Vetterli et al. 1989]{Vetterli89}Vetterli, M.C., H$\odot{a}$usser, O., Abegg, R.,
Alford, W.P., Celler, A., Frekers, D., Helmer, R., Henderson, R.,
Hicks, k.H., Jackson, K.p., Jappesen, R.G., Miller, C.A, Raywood,
K., Yen, S. 1989, Phys. Rev. C, \textbf{40}, 559
\bibitem[Wallace et al. 1981]{Wallace81} Wallace, R.K., Woosley, S.E. 1981, Astrophys. J. Suppl., \textbf{45}, 389-420
\bibitem[Wang et al. 1988]{Wang88} Wang, D., Rapaport, J., Horen, D.J., Brown, B.A., Gaarde, C., Goodman, C.D., Sugarbaker, E., Taddeucci,
T.N. 1988, Nucl. Phys. A, 480, 285

%

\end{thebibliography}
\end{document}